\documentclass[journal=jacsat,manuscript=article,pdftex]{achemso}
\SectionNumbersOn
\usepackage[whole]{bxcjkjatype}
\usepackage{bm}
\usepackage{url}
\usepackage{ulem}
\usepackage{xcolor}
\usepackage{comment}
\usepackage{amsmath}
\usepackage{amssymb}
\usepackage{txfonts}
\usepackage{dcolumn}
\usepackage{tabularx}
\usepackage{multirow}
\usepackage{graphicx}
\usepackage{arydshln}
\newcommand{\tcom}[1]{{}} 
\newcommand{\tadd}[1]{{\color{black}#1}}
\newcommand{\trem}[1]{{}}
 
\newcommand{\ghost}[1]{{}}
\renewcommand{\thefootnote}{\alph{footnote}}

\newcommand{\pbbody}{Pb$_2$Ti$_4$O$_9$F$_2$}

\newcommand{\bibody}{Bi$_2$Ti$_4$O$_{11}$}

\title{\tadd{Anionic Ordering in {\pbbody} Revisited by Nuclear Magnetic Resonance and Density Functional Theory}}
\author{Kengo Oka}
\affiliation[kindai]{Department of Applied Chemistry, Faculty of Science and Engineering, Kindai University, Higashiosaka, Osaka 577-8502, Japan}
\alsoaffiliation[equal]{These two authors contributed equally.}
\email{koka@apch.kindai.ac.jp}
\author{Tom Ichibha}
\affiliation[ornl]{Materials Science and Technology Division, Oak Ridge National Laboratory, Oak Ridge, TN 37831, USA}
\alsoaffiliation[equal]{These two authors contributed equally.}
\email{ichibha@icloud.com}
\author{Daichi Kato}
\affiliation[kyodai]{Department of Energy and Hydrocarbon Chemistry, Graduate School of Engineering, Kyoto University, Kyoto 615-8510, Japan}
\author{Yasuto Noda}
\affiliation[kyodai2]{Division of Chemistry, Graduate School of Science, Kyoto University, 606-8502 Kyoto, Japan}
\author{Yusuke Tominaga}
\affiliation[kyodai2]{Division of Chemistry, Graduate School of Science, Kyoto University, 606-8502 Kyoto, Japan}
\author{Kosei Yamada}
\affiliation[kindai]{Department of Applied Chemistry, Faculty of Science and Engineering, Kindai University, Higashiosaka, Osaka 577-8502, Japan}
\author{Mitsunobu Iwasaki}
\affiliation[kindai]{Department of Applied Chemistry, Faculty of Science and Engineering, Kindai University, Higashiosaka, Osaka 577-8502, Japan}
\author{Naoki Noma}
\affiliation[joint]{Joint Research Center, Kindai University, Higashiosaka, Osaka 577-8502, Japan}
\author{Kenta Hongo}
\affiliation[jaist1]{Research Center for Advanced Computing Infrastructure, JAIST, Asahidai 1-1, Nomi, Ishikawa 923-1292, Japan}
\author{Ryo Maezono}
\affiliation[jaist2]{School of Information Science, JAIST, Asahidai 1-1, Nomi, Ishikawa 923-1292, Japan}
\author{Fernando A. Reboredo}
\affiliation[ornl]{Materials Science and Technology Division, Oak Ridge National Laboratory, Oak Ridge, TN 37831, USA}

\newcommand\blfootnote[1]{%
  \begingroup
  \renewcommand\thefootnote{}\footnote{#1}%
  \addtocounter{footnote}{-1}%
  \endgroup
}
\begin{document}
\maketitle
\newpage
\begin{abstract}
  \tadd{
    A combination of $^{19}$F magic angle spinning (MAS) nuclear magnetic resonance (NMR) and density functional theory (DFT) were used to study the ordering of F atoms in {\pbbody}. 
    This analysis revealed that F atoms predominantly occupy two of the six available inequivalent sites in a ratio of 73:27.
    DFT-based calculations explained the preference of F occupation on these sites and quantitatively reproduced the experimental 
    occupation ratio, independent of the choice of functional. 
    We concluded that the Pb atom's 6$s^2$ lone pair may play a role ($\sim$0.1 eV/f.u.) in determining the majority and minority F occupation sites
    with partial density of states and crystal orbital Hamiltonian population analyses applied to the DFT wave functions.
  }
\end{abstract}
\blfootnote{This manuscript has been authored by UT-Battelle, LLC, under contract DE-AC05-00OR22725 with the US Department of Energy (DOE). The US government retains and the publisher, by accepting the article for publication, acknowledges that the US government retains a nonexclusive, paid-up, irrevocable, worldwide license to publish or reproduce the published form of this manuscript, or allow others to do so, for US government purposes. DOE will provide public access to these results of federally sponsored research in accordance with the DOE Public Access Plan (http://energy.gov/downloads/doe-public-access-plan).}
\section{Introduction}
\label{sec.intro}\ghost{sec.intro}
Understanding the mechanisms that produce ionic ordering in materials may lead to 
control of the self-assembly of ordered super lattices on an underlying crystal structure.
Because ordered and disordered structures have remarkably different electronic
and transport properties, 
controlling ordering is a key step in material design.
For example,
the properties of mixed-anion compounds strongly depend on the degree of order
or disorder of the anions\cite{2019HK_YK,2022KM_HK}.
Anionic ordering can cause heteroleptic coordinations or low-dimensional structures,
which in turn modify electronic properties\cite{2019HK_YK,2022KM_HK}.
This intriguing possibility to modify a material has attracted significant attention
\cite{2018KAG,2018KUR_AST,2018OSH_UND,2020OSH_TWO}.
\tadd{Some ABX$_3$ perovskites are typical examples.}
SrTaO$_2$N and BaTaO$_2$N have a high dielectric constant 
because of their O/N anionic ordering\cite{2004KIM_CHA,2007PAG_LOC,2010YAN_ANI}. 
In addition, an oxyhydride SrVO$_2$H shows two-dimensional 
electron conduction and compression anisotropy because of the O/H anionic ordering\cite{2017YAM_THE}.

\vspace{2mm}
\tadd{In complex composite materials,} multiple anions and cations are heteroleptically coordinated, and
their concentrations typically obey valence charge neutrality conditions. 
When differences in ionic radii, electronegativity, or polarizability are large,
\tadd{some} materials tend to exhibit ionic ordering\cite{2013JPA,2006AF}. 
However, although O$^{2-}$ and F$^{-}$ are neighboring anions in the periodic table,
the oxyfluorides nevertheless exhibit anionic ordering, depending on the structure.
\tadd{Whereas} simple cubic perovskites 
(SrFeO$_2$F\cite{2014THO}, 
BaFeO$_2$F\cite{2011BER}, 
PbScO$_2$F\cite{2008KAT}, 
BaScO$_2$F\cite{1998NEE}, 
AgFeOF$_2$\cite{2018TAK}, 
BaInO$_2$F\cite{2019KAT}, 
and AgTiO$_2$F\cite{2020INA}) 
adopt disordered configurations, 
a variety of Ruddlesden--Popper-type layered perovskites 
(Sr$_2$CuO$_2$F$_2$\cite{1994AI}, 
Sr$_2$FeO$_3$F\cite{1999SIM,2001HEC}, 
Ba$_2$InO$_3$F\cite{1995NEE}, 
Ba$_2$ScO$_3$F\cite{1996NEE}, 
Sr$_2$MnO$_3$F\cite{2016SU}, 
Sr$_3$Fe$_2$O$_{5-x}$F$_y$\cite{1999SIM,2011TSU}) 
exhibit ordered configurations of F$^-$\cite{1998DU}.
\tadd{
  For an ordered structure to be formed,
  migration energy barriers must be small enough
  to allow the minimum free energy configuration
  to be achieved via practical annealing temperatures and times.
}
\tadd{The lowest energy structure} frequently can be found
by considering Pauling's second rule \cite{1960LP},
\tadd{
  which explains that an ionic structure will be stable when the sum
  of the strength of the electrostatic bonds around an ion are equal to its charge.
}
Thus, \tadd{according to Pauling's second rule}, F$^-$ \tadd{should} 
prefer a more open site compared with O$^{2-}$, leading to O/F anionic ordering.
\tadd{Some nonlayered oxyfluorides with or akin to perovskite structure} 
showing \tadd{a complete or partial} anionic ordering \tadd{have been reported}, such as Pb$_2$Ti$_2$O$_{5.4}$F$_{1.2}$\cite{2016OKA}
Pb\tadd{$_2$}OF$_2$ \cite{2019INA}, and Pb$_2$Ti$_4$O$_9$F$_2$ \cite{2015OKA}.
\tadd{The anionic ordering of some nonlayered oxyfluorides} is considered to be due to the Jahn--Teller distortion
by the 6$s^2$ lone pair \cite{2016OKA,2015OKA}.

\vspace{2mm}
\tadd{
  Among the nonlayered oxyfluorides with anionic ordering, {\pbbody} is especially attractive to study
  because this material uniquely has an isostructural oxide {\bibody}\cite{2015OKA, 1995KAH}.
}
Both Pb$^{2+}$ and Bi$^{3+}$ have the same electronic configuration and 6$s^2$ lone pairs.
Bi$_2$Ti$_4$O$_{11}$ undergoes antiferroelectric-paraelectric transitions from $C2/c$ to $C2/m$,
whereas Pb$_2$Ti$_4$O$_9$F$_2$ does not\cite{2015OKA,1995KAH}.
The high-temperature paraelectric phase of Bi$_2$Ti$_4$O$_{11}$~($C2/m$) adopts
the same space group symmetry as Pb$_2$Ti$_4$O$_9$F$_2$\cite{1995KAH}.
\tadd{
  Because the {\bibody}-type structure is significantly low symmetric,
  an anionic ordering can exist in {\pbbody}.
  Structural analyses based on the synchrotron X-ray diffraction (SXRD) patterns revealed the presence of anionic ordering \cite{2015OKA}. 
  However, this conclusion is debatable 
  because heavier Pb atoms in the system could
  hamper the identification of F occupation sites. 
}

\vspace{2mm}
\tadd{
  In this work, 
  the anionic ordering in {\pbbody} was reexamined via a combination of $^{19}$F magic angle spinning (MAS) nuclear magnetic resonance (NMR) 
  experiments and density functional theory (DFT) simulations. The $^{19}$F MAS NMR analysis revealed that F atoms randomly occupy 
  two of the six sites in a ratio of 73:27, overturning the previous conclusion that F atoms selectively occupy a single site 
  \cite{2015OKA}.
  DFT calculations identified the majority and minority F occupation sites and quantitatively reproduced the experimental
  occupation ratio, independent of the choice of functional.
  Partial density of states (PDOS) and crystal orbital Hamiltonian population (COHP) analyses were performed on the DFT results, showing that
  the 6$s^2$ lone pairs may play a role ($\sim$0.1 eV/f.u.) in determining the majority and minority F occupation sites.
  On the other hand, DFT calculations revealed that the low-symmetric anionic coordinates
  around the cations may barely be due to the the steric effects of 6$s^2$ lone pairs. This result goes
  against the current discussion in this class of materials and
  implies that the influence of 6$s^2$ lone pairs on the structural distortion might 
  be similarly not significant in some of the other Pb--based oxyfluorides
  such as Pb$_2$Ti$_2$O$_{5.4}$F$_{1.2}$\cite{2016OKA} and Pb$_2$OF$_2$\cite{2019INA}.
}

\section{Experimental details}
\label{sec.methods}\ghost{sec.methods}
The powder samples of {\pbbody} and {\bibody} were prepared by solid-state reaction,
as previously reported\cite{2015OKA,1995KAH}.
The Pb$_2$Ti$_4$O$_9$F$_2$ was synthesized from 
a mixture of PbO (99.9~\%, Rare Metallic Co.), 
PbO$_2$ (99.9~\%, Rare Metallic Co.), 
PbF$_2$ (99.9~\%, Rare Metallic Co.), 
and TiO$_2$ (rutile, 99.9~\%, Rare Metallic Co.)
powders that were weighed to be 10~mol \% F-rich to compensate for the loss of F
during the reaction\cite{2012KAT}. The pelletized mixture was sealed in an evacuated Pyrex tube
and treated at 823 K for 12 h \tadd {in an electric furnace, followed by natural cooling to room temperature.}
The Bi$_2$Ti$_4$O$_{11}$ was synthesized
from a stoichiometric mixture of Bi$_2$O$_3$ (99.9\%, Rare Metallic Co.) and
TiO$_2$ (99.9\%, Rare Metallic Co.) powders. The pelletized mixture was treated at 1273 K
for 12 h in air in an electric \tadd{furnace}, 
\tadd {followed by natural cooling  to room temperature}.

\vspace{2mm}
\tadd{
  Production of a single phase for both {\pbbody} and {\bibody} samples 
  were confirmed via SXRD.
}
SXRD patterns were collected with a large Debye–Scherrer camera 
installed at beamline BL02B2 of the Super Photon ring-8 Gev (SPring-8) synchrotron
radiation facility
using a glass capillary and a solid-state detector \cite{2017KAW}. 
The crystallographic parameters were 
refined by the Rietveld method using the RIETAN-FP program\cite{2007IZU}.
The electron density distributions were estimated by the maximum entropy method (MEM) 
using the Dysnomia program\cite{2013MOM}.

\vspace{2mm}
\tadd{
  Solid-state NMR experiments were conducted on a homemade spectrometer with a 4 mm T3 probe (Varian) in a magnetic field of 4.7 T.
  All $^{19}$F NMR transients under MAS were accumulated using a background suppression method. 
  The radio frequency field strength was 100 kHz, corresponding to 2.5~{\textmu}s of $\pi/2$ pulse length.
  The longitudinal relaxation time ($T_1$) was obtained by analyzing a build-up curve measured with a saturation recovery method.
  Rotational resonance experiments, which allow for solving whether  
  the chemical-shift filter was set to half of the inverse of the difference between two signals and
  the MAS rate, was set to the inverse of the difference between two signals
  in rotational resonance experiments (further details given in supporting information [SI]).
}

\section{Calculation details}
\label{sec.calc}\ghost{sec.calc}
The DFT calculations were performed with Quantum Espresso\cite{2009GIA}.
\tadd{Perdew--Burke--Ernzerhof (PBE)\cite{PBE}, Becke--Lee--Yang--Parr (BLYP)\cite{BLYP1,BLYP2}, and Perdew--Wang 1991 (PW91)\cite{PW911,PW912}} semi-local
exchange--correlation functionals were employed. 
The valence orbitals were expanded with plane waves. 
The cutoff energy was 100 Ry, and the $k$-point mesh was 7$\times$7$\times$5 for a unit cell. 
\tadd{
  With this choice of parameters, the energy difference between
  the first and second most stable anionic ordering patterns of {\pbbody} converged below 2 meV/f.u.
  These two ordering patterns are denoted as F-in-site6 and F-in-site5 in Section \ref{sec.nmr}.
}
The core orbitals were described by the projector augmented wave (PAW) method\cite{Kresse1999}.
PAW pseudopotentials were taken from the pslibrary\cite{2014ADC}.
Comparing the results of PAW and ultrasoft pseudopotentials in the pslibrary\cite{2014ADC} 
revealed that the relative total energies among the different anionic orderings are identical within 1~meV/f.u.
Therefore, the errors from pseudopotential approximation would be negligible for this system. 
Ultrasoft pseudopotentials were used to calculate the electrostatic energies because PAW pseudopotentials cannot separately provide the electrostatic energies
because the one-center term includes both electrostatic and exchange-correlation energies\cite{1994PEB}.
\tadd{    
  The PDOS of {\pbbody} was obtained with the PBE functional.
  The LOBSTER code \cite{2020NEL} was used to perform 
  the COHP analyses based on the PBE-DFT results with the pbeVaspFit2015 basis set \cite{2013MAI,2016MAI}.
  Every DFT calculation was performed for the energetically optimized structure.
}

\section{Results and discussion}
\subsection{\tadd{Determination of anionic configurations}}
\label{sec.nmr} 
\begin{figure}[htbp]
  \centering
  \includegraphics[width=0.5\hsize]{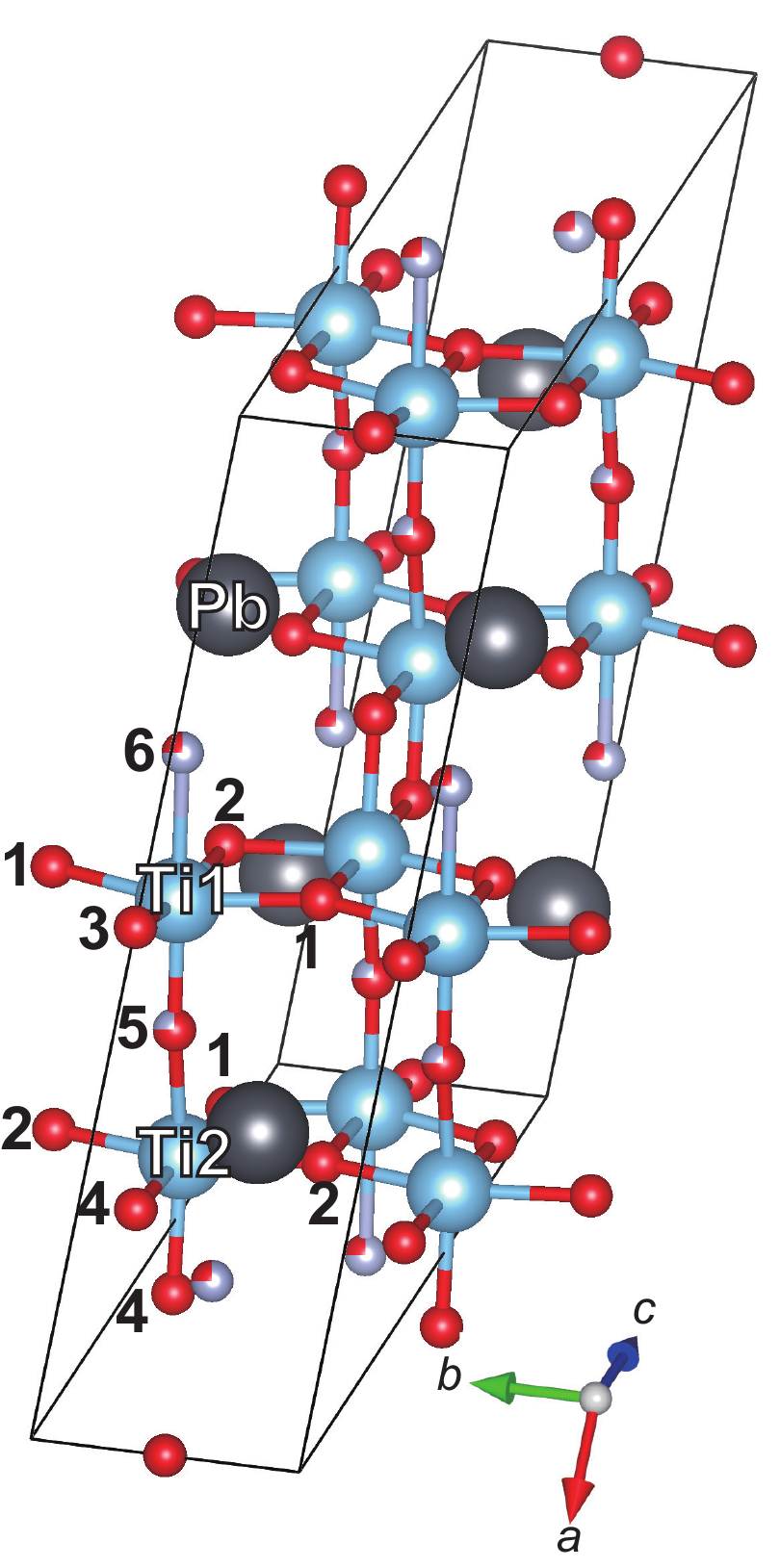}
  \caption{
    \tadd{
      \label{fig.sites}\ghost{fig.sites}
      The crystal structure of {\pbbody} at 300~K.
      The numbers indicate the numbers of anion sites.
    }
  }      
\end{figure}

\ghost{sec.nmr}
\tadd{
  As shown in Figure \ref{fig.sites}, {\pbbody} has six different anion sites, 
  which are denoted by site 1--6. 
  Previous structural analysis based on SXRD patterns has indicated that F atoms selectively occupy site 6, 
  which is the closest site to the Pb atom\cite{2015OKA}. 
  Here, that earlier conclusion is reexamined using NMR experiments and DFT simulations. 
}

\vspace{2mm}
\tadd{
  A $^{19}$F MAS NMR spectrum of Pb$_2$Ti$_4$O$_9$F$_2$ is shown together with a peak fitting result in Figure~\ref{fig.nmr}.
  By fitting the spectrum with three Gaussians, the peak positions were obtained as $-45$~ppm, $-58$~ppm, and $-63$~ppm, with an area ratio of 26.1:1.8:72.1.
  The peak positions correspond to the $^{19}$F atoms in different distinct sites, and the area ratio indicates their occupancy ratio.
  Through-space correlation NMR experiments were conducted to confirm whether the $^{19}$F atoms showing in the two main peaks exist in the same crystal phase.
  When the MAS rate was matched to the resonant frequency difference of the two main peaks, both peak intensities varied periodically with mixing time.
  However, when the MAS rate was faster than the resonant frequency difference, neither peak intensity changed (details in SI).
  This means that the dipole interaction that disappeared because of MAS was reintroduced by
  rotational resonance, and the magnetization was exchanged
  during the mixing time.
  Thus, the F sites showing the two main signals are nearby in the same crystalline phase, Pb$_2$Ti$_4$O$_9$F$_2$.
  \begin{figure}[ttbp]
    \centering
    \includegraphics[width=\hsize]{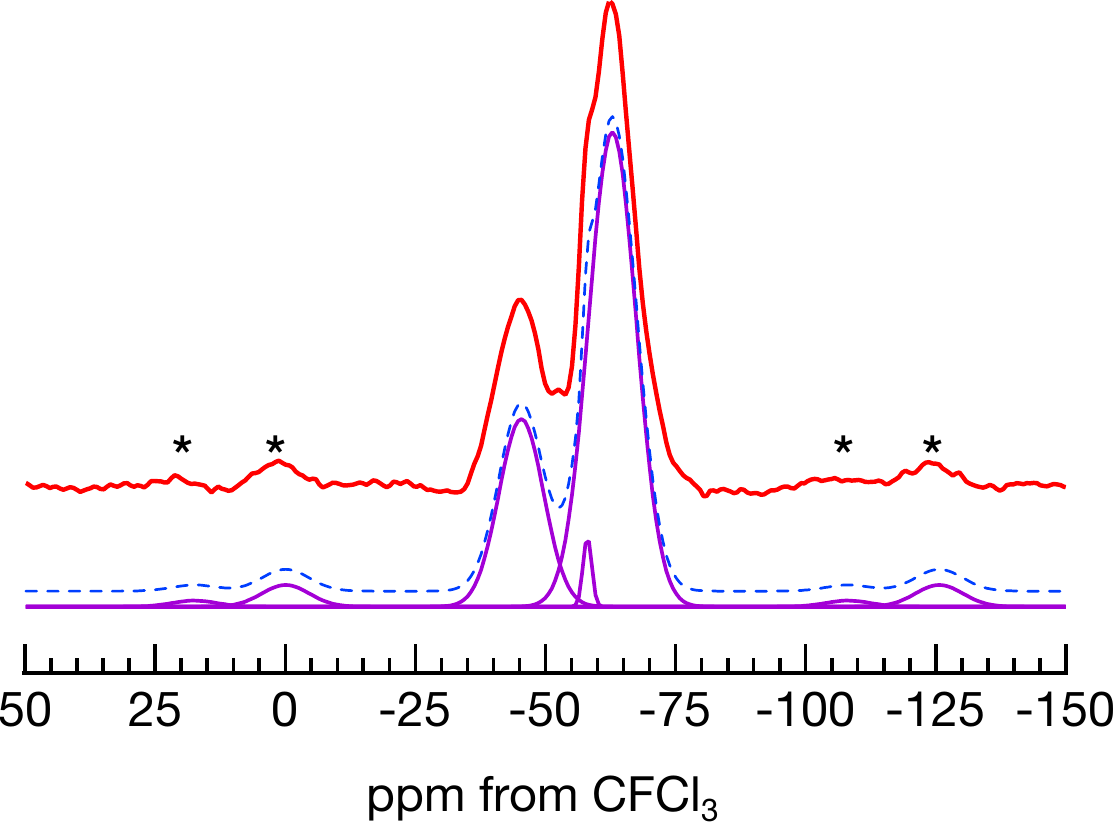}
    \caption{
      \label{fig.nmr}
      \tadd{
        A $^{19}$F MAS NMR spectrum of Pb$_2$Ti$_4$O$_9$F$_2$ (top) and peak fitting result (bottom). The asterisks indicate spinning sidebands.
      }
    }      
  \end{figure}
}

\vspace{2mm}
\tadd{  
  The position of the F occupation sites 
    corresponding to the two main peaks in the $^{19}$F MAS NMR experiment must be determined using another method.
  Therefore, 
  the energies of different F occupation patterns were evaluated with DFT 
  to find the majority and minority F occupation sites corresponding to the two main peaks.
  In reality, numerous F occupation patterns are possible, and we cannot study all of them.
  Therefore, targets were restricted to be the cases in which F atoms selectively occupy each of the sites, and 
  the F atom's stability in each site was evaluated.  
  Table \ref{tab.TotalEnergy} lists the relative total energies given by PBE \cite{PBE}, 
  BLYP \cite{BLYP1,BLYP2}, and PW91 \cite{PW911,PW912} functionals.
  In the table, for example, F-in-site2 indicates the case in which F atoms selectively occupy site~2.
  The F stability at site~3 was evaluated as
  $E$(F-in-site3) $\equiv$ 2$E$(one F atom in site~3 and the other in site~6 in the unit cell) $-$ $E$(F-in-site6)
  because the multiplicity of site~3 is one (multiplicity of the other sites is two).
  The three functionals qualitatively and quantitatively agree with each other. 
  F-in-site6 and F-in-site5 give the first and second lowest energies, so 
  the majority and minority F occupation sites are sites~6 and 5, respectively. 
  The experimental third tiny peak at $-$58 ppm corresponds to the F occupancy in 
  site~4.
  The percentages in the parentheses in Table~\ref{tab.TotalEnergy} indicate the ratios 
  of Boltzmann factors of the relative energies under the synthesis temperature, 823 K.
  These percentages correspond to the F occupation ratio of the anion sites. 
  The ratio between the percentages of F-in-site6 and F-in-site5, approximately 70:30, closely agrees with that 
  between the experimental F occupation ratios of the majority and minority sites, 73:27.
}

\begin{table}[htbp]
  \begin{center}
    \caption{
      \label{tab.TotalEnergy}\ghost{tab.TotalEnergy}      
      \tadd{
        Comparisons of the energy differences (eV/f.u.)
        between alternative anionic ordering patterns in {\pbbody}.
        All energies are differences with the energy of F-in-site6. 
        The percentages indicate the ratios of the corresponding Boltzmann
        factors at the synthesis temperature, 823 K.
        F-in-site3 does not indicate an anionic ordering pattern
        different from the others, as described in the main text.
      }
    }
    \begin{tabular}{ccccccc}
      \hline
      & \multicolumn{2}{c}{$\Delta E$ (PBE)} & \multicolumn{2}{c}{$\Delta E$ (BLYP)} & \multicolumn{2}{c}{$\Delta E$ (PW91)} \\
      \hline
      F-in-site1 & 1.019 &  (0.0\%) & 0.965 &  (0.0\%)  & 1.018 &  (0.0\%) \\
      F-in-site2 & 0.935 &  (0.0\%) & 0.889 &  (0.0\%)  & 0.937 &  (0.0\%) \\
      F-in-site3 & 0.513 &  (0.1\%) & 0.514 &  (0.1\%)  & 0.543 &  (0.0\%) \\
      F-in-site4 & 0.321 &  (0.8\%) & 0.308 &  (0.9\%)  & 0.322 &  (0.7\%) \\
      F-in-site5 & 0.061 & (29.5\%) & 0.066 & (28.0\%)  & 0.058 & (30.3\%) \\
      F-in-site6 & 0.000 & (69.7\%) & 0.000 & (71.1\%)  & 0.000 & (69.0\%) \\
      \hline
    \end{tabular}
  \end{center}
\end{table}

\subsection{\tadd{Reason why sites 5 and 6 have a preference for F atoms}}
\label{sec.results2}
\begin{figure*}[htbp]
  \centering
  \includegraphics[width=\hsize]{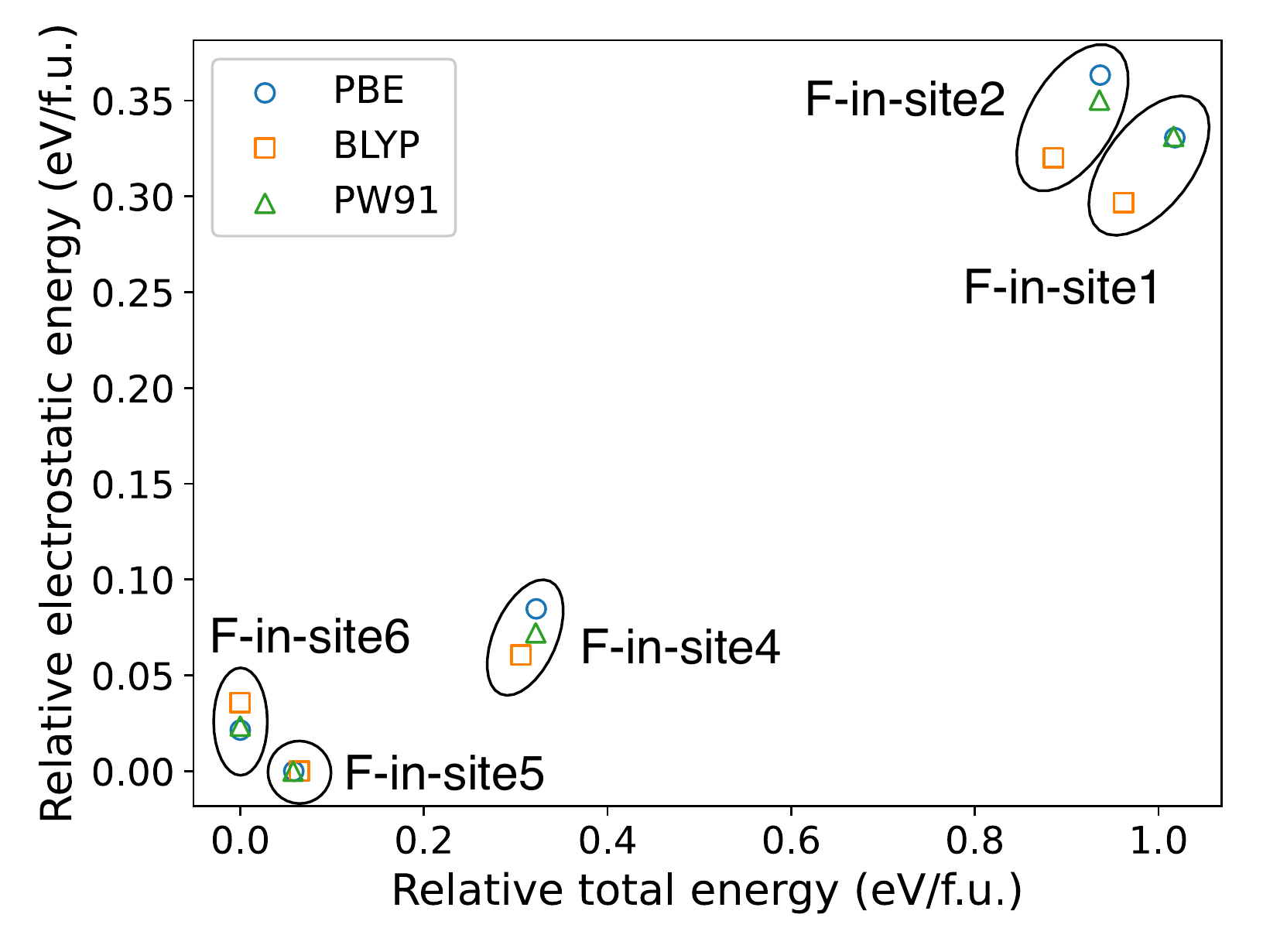}
  \caption{
    \label{fig.static}\ghost{fig.static}
    Comparison of electrostatic energies and total energies calculated by DFT,
    using \tadd{PBE (blue circles), BLYP (orange squares), and PW91 (green triangles) functionals}.
    \tadd{
      The minimum electrostatic energy (i.e., F-in-site5) and total energy (i.e., F-in-site6)
      are set to be zero, and the others are relative to this for every functional.
    }
  }      
\end{figure*}

\ghost{sec.results2}
To understand the origin of the F-in-site6 \tadd{and F-in-site5} stabilization, 
\tadd{PBE\cite{PBE}, BLYP\cite{BLYP1,BLYP2}, and PW91\cite{PW911,PW912}} functionals
were used to compare the electrostatic energies with the total energies, as shown
in Figure~\ref{fig.static}. 
The results of \tadd{the three} functionals are qualitatively consistent with each other.
\tadd{The total energies in Figure~\ref{fig.static} are identical to those listed in 
Table~\ref{tab.TotalEnergy}.}
The total energies are roughly proportional to the electrostatic energies, 
consistent with a previous systematic study on NdNiO$_2$F by simulations \cite{2019KUR}. 
However, F-in-site6, which had \tadd{the lowest total energy}, is an exception to Pauling's second rule \cite{1960LP}
\tadd{because} the \tadd{lowest} electrostatic energy \tadd{was predicted for the F-in-site5.}

\begin{figure}[htbp]
  \centering
  \includegraphics[width=\hsize]{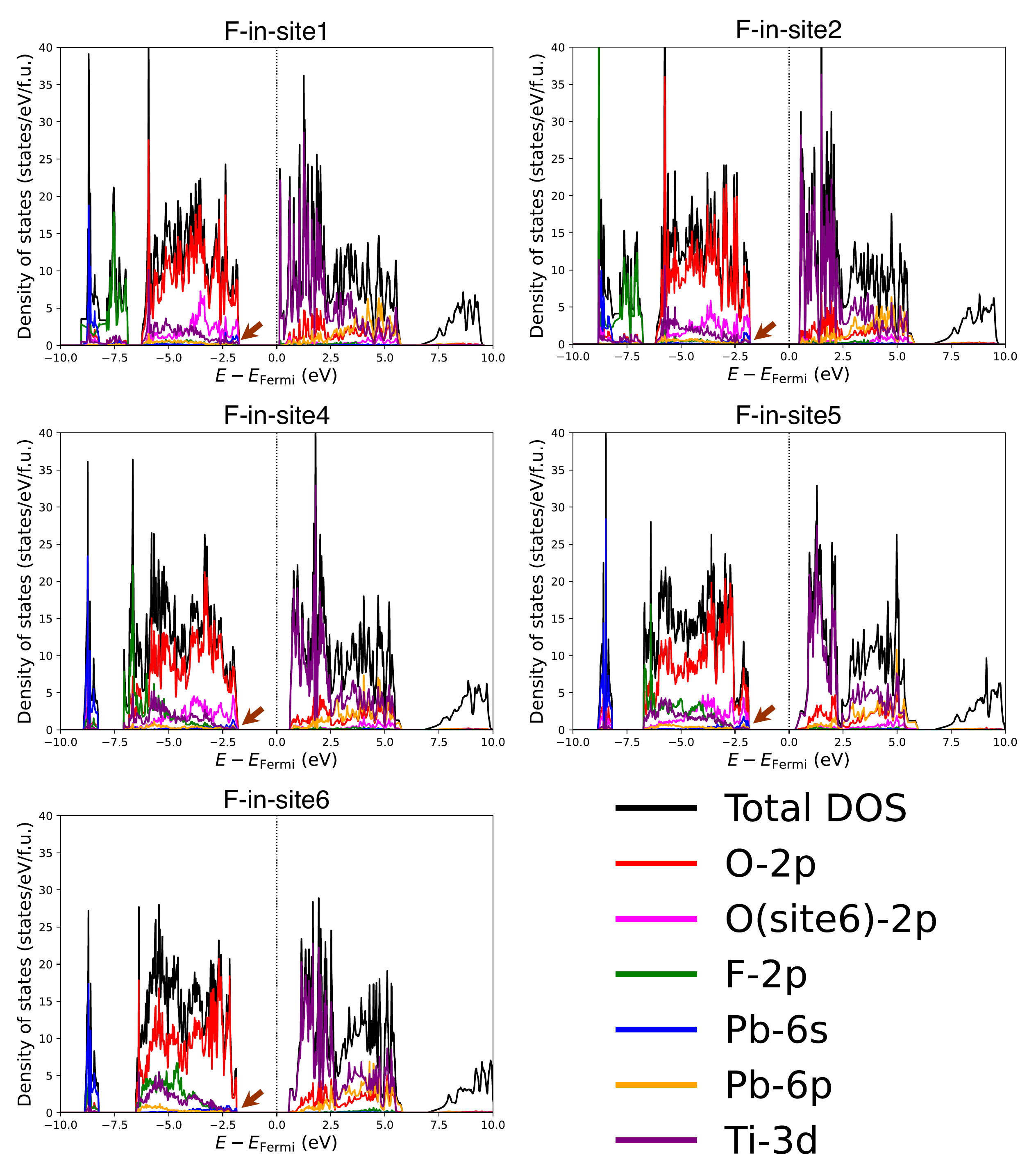}
  \caption{
    \label{fig.pdos}\ghost{fig.pdos}
    PDOS of different F occupation patterns obtained by DFT.
    The arrows indicate the peaks accompanied by the antibonding
    hybrid orbitals explained by the lone-pair model.
    For the F-in-site6 case, the O(site6)-2$p$ distribution is not shown
    because site 6 is occupied by F atoms.
  }      
\end{figure}

\vspace{2mm}
The
reason of the small deviation ($\sim$0.1 eV/f.u.) from Pauling's second rule \cite{1960LP}
is unclear.  
However, the steric effects introduced by the 6$s^2$ lone pairs may explain this deviation. 
Figure~\ref{fig.pdos} shows the PDOS of each F ordering pattern in {\pbbody} \tadd{given by the PBE functional}. 
A peak of Pb-6$s$ (indicated by an arrow) hybridized with O-2$p$ exists at the valence band maximum (VBM). 
This peak is considered to be a signal of the orbital hybridization 
described by the revised lone-pair (RLP) model\cite{2011WAL}. 
This model proposes that the antibonding hybridized orbital of 
Pb-6$s$ and O-2$p_{z}$ is further stabilized by hybridizing with the Pb-6$p$ orbital.
This antibonding orbital distributes opposite to the O \cite{2011WAL}
and interrupts bonding with the other anions.
The PDOS figures show that F-in-site6 has the smallest Pb-6$s$ peak (i.e.,~the weakest RLP hybridization),
presumably owing to the absence of O6 (i.e., an O ion in site~6) that is the closest to Pb and has the largest 2$p$ \tadd{density of states (DOS)} at the VBM.
\tadd{
  Furthermore, the magnitudes of the hybridization between Pb-6$s$ and O/F6-2$p$
  were evaluated for F-in-site5 and F-in-site6 in terms of integrated COHP (ICOHP) by the COHP analysis applied to the PBE--DFT results.
  Here, ICOHP indicates how much the hybridization contributes to the stabilization by the binding.
  The ICOHPs of Pb-O6 in F-in-site5 and Pb-F6 in F-in-site6 are $-$2.81 and $-$0.90~eV,
  so the contribution of the 6$s^2$ lone pair is certainly smaller when F atoms occupy
  site~6.
}
This explanation for the F-in-site6 stabilization is an exception to Pauling's second rule \cite{1960LP}.

\subsection{Comparison between $\bm{\mathrm{Pb_2Ti_4O_9F_2}}$ and $\bm{\mathrm{Bi_{2}Ti_{4}O_{11}}}$}
\label{sec.results1}
\begin{figure*}[htbp]
  \centering
      {\Large {\pbbody}}\\
      \begin{minipage}{0.32\hsize}
        \includegraphics[width=\hsize]{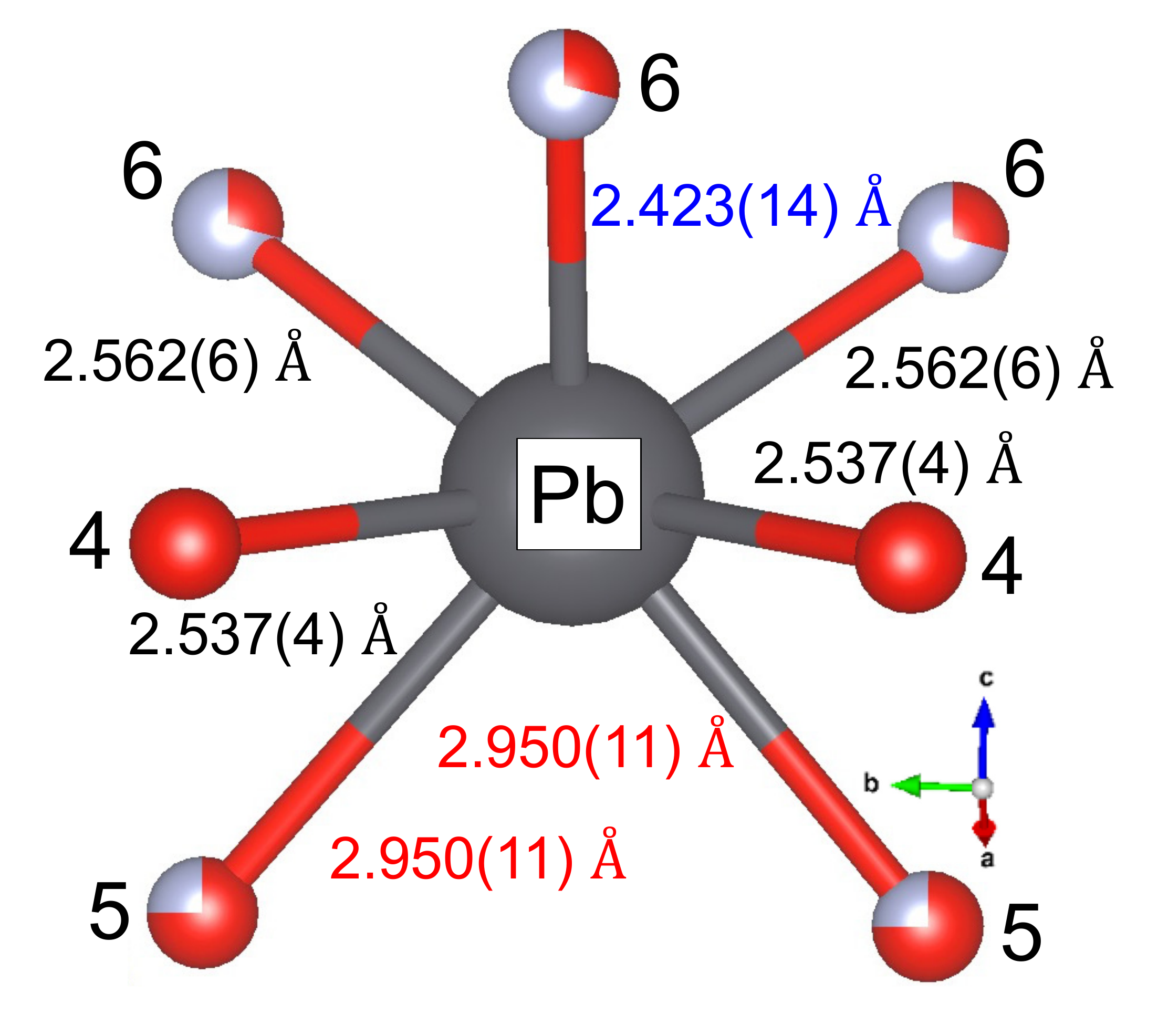}
      \end{minipage}
      \begin{minipage}{0.32\hsize}
        \includegraphics[width=\hsize]{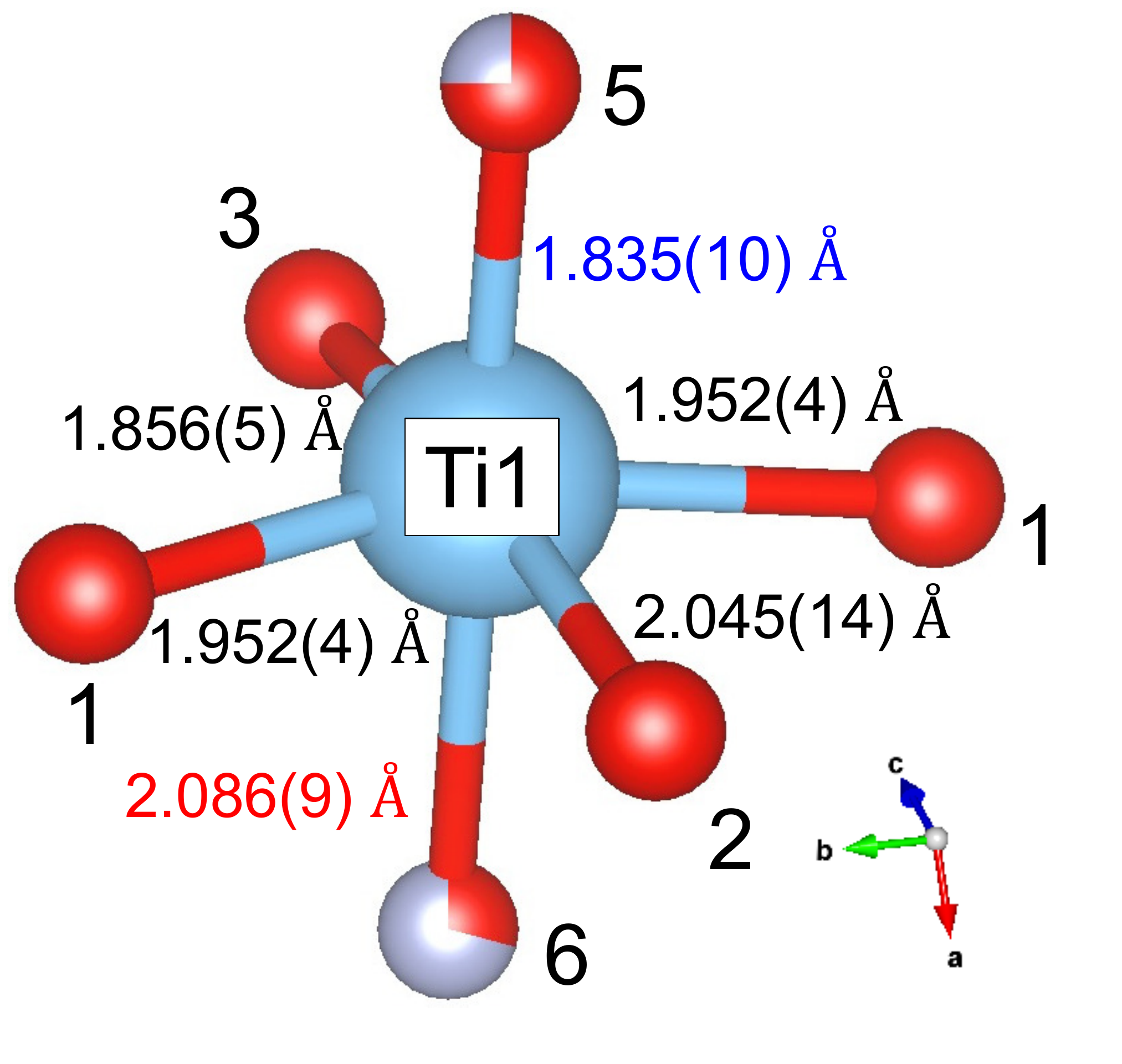}
      \end{minipage}
      \begin{minipage}{0.32\hsize}
        \includegraphics[width=\hsize]{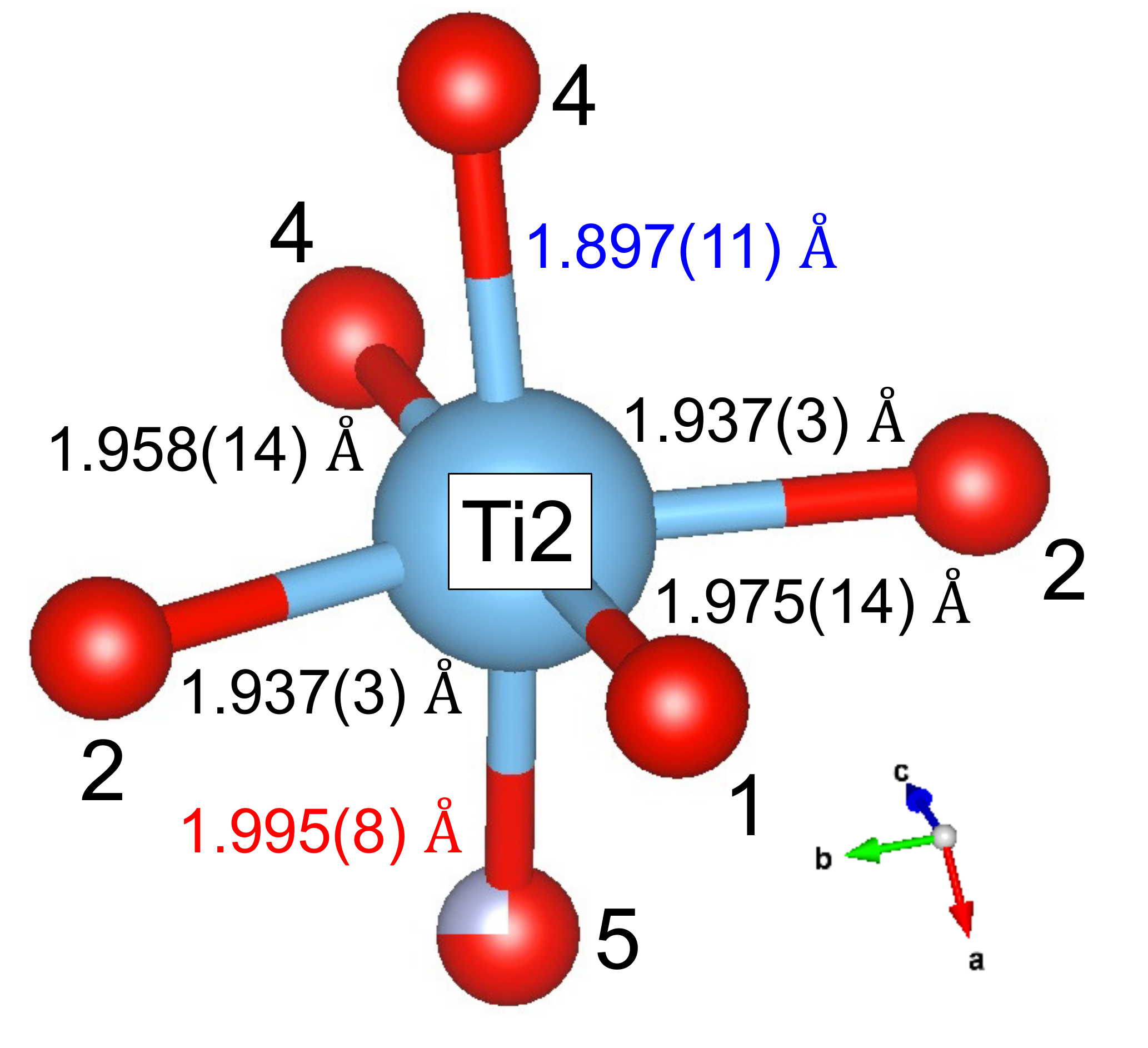}\\
      \end{minipage}
          {\Large {\bibody}}\\
          \begin{minipage}{0.32\hsize}            
            \includegraphics[width=\hsize]{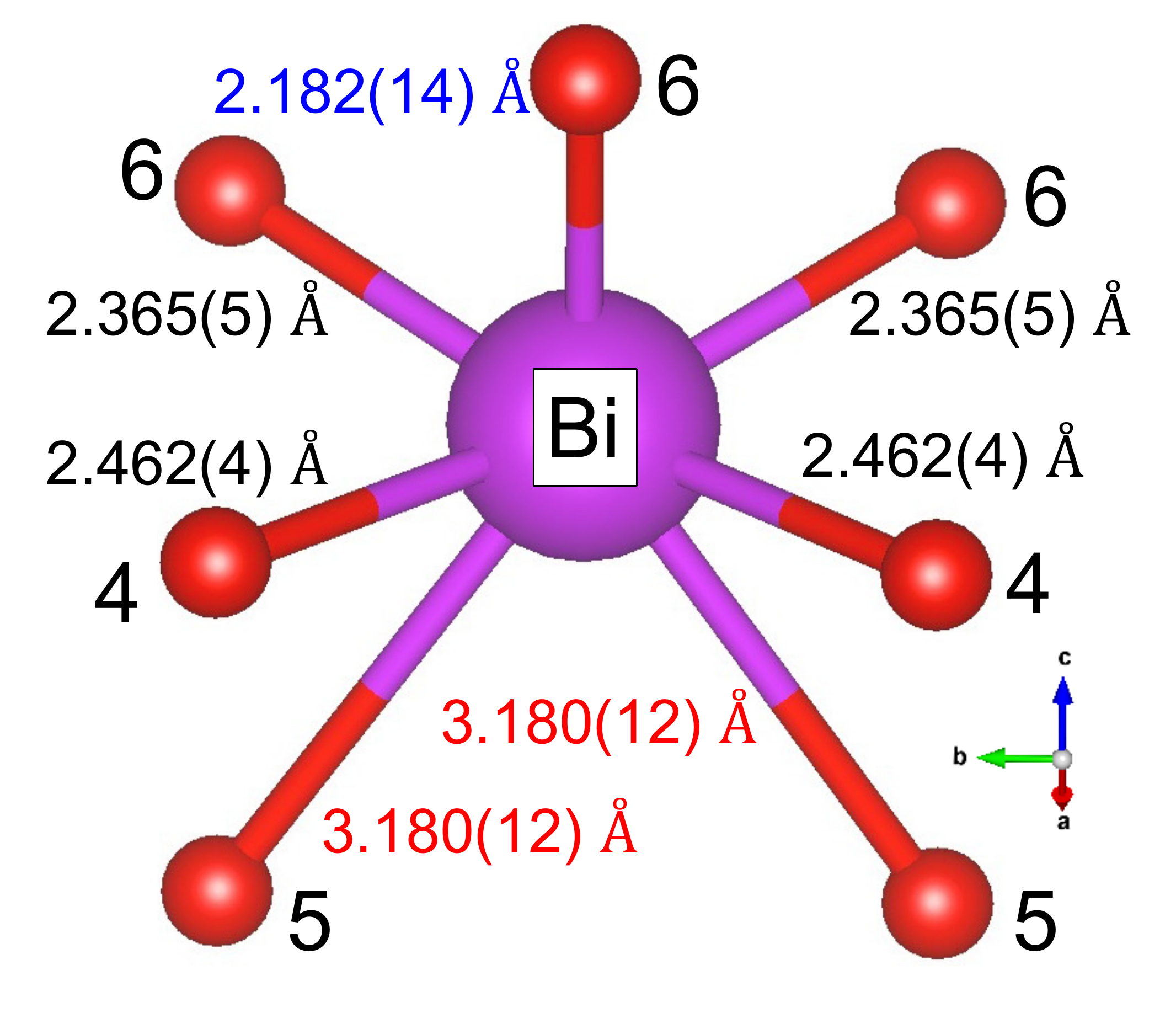}
          \end{minipage}
          \begin{minipage}{0.30\hsize}
            \includegraphics[width=\hsize]{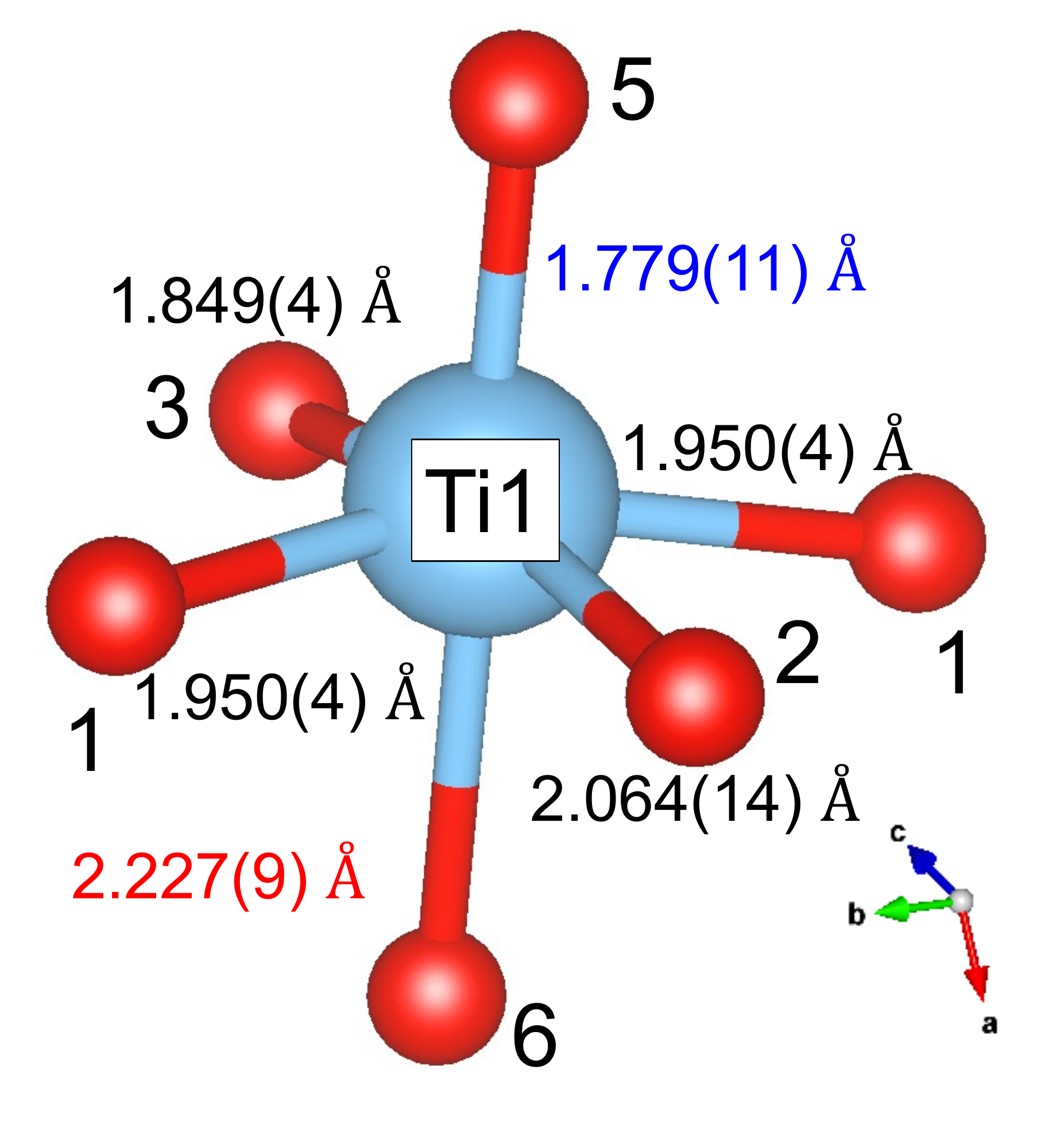}
          \end{minipage}
          \begin{minipage}{0.32\hsize}
            \includegraphics[width=\hsize]{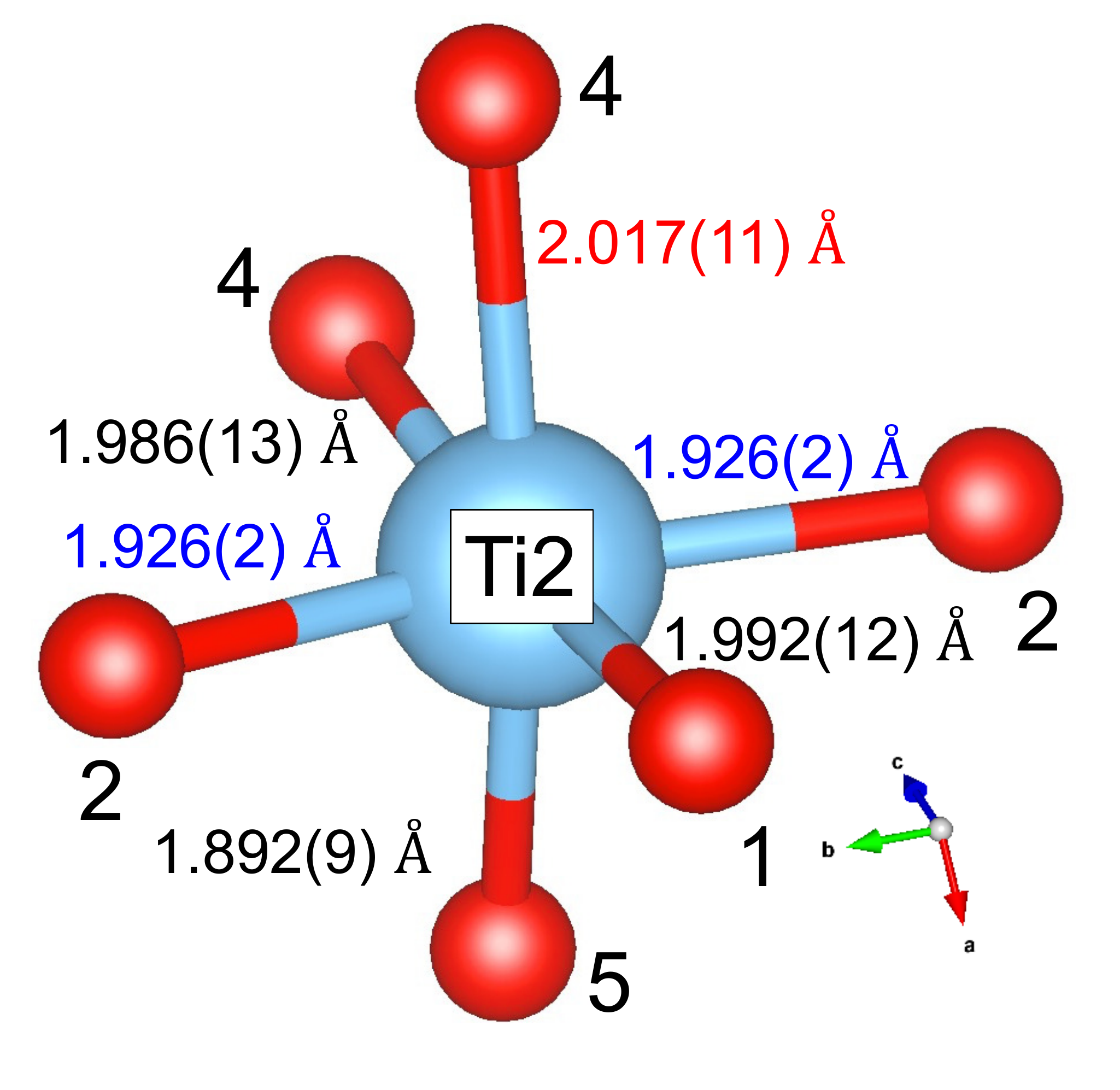}
          \end{minipage}
          \caption{
            \label{fig.strExpt}\ghost{fig.strExpt}
            \tadd{
              Experimental geometries of {\pbbody} and {\bibody}
              obtained via Rietveld analysis of SXRD patterns.
              The smallest and largest interatomic distances are
              shown with blue and red colors, respectively.
            }
          }      
\end{figure*}

\ghost{sec.results1}
\tadd{
  Figure \ref{fig.strExpt} compares the experimental structures of {\pbbody} and {\bibody},
  which were obtained by the \hbox{Rietveld} analysis of SXRD patterns.
  The SXRD patterns are provided in the SI.
  Comparing the isostructural {\pbbody} and {\bibody},
  the anionic configurations around the cations are more symmetric in {\pbbody} than in {\bibody}. 
  For example, the difference of the smallest (blue) and largest (red) interatomic distances are smaller in {\pbbody} than in {\bibody}.
  It is worth noting that this trend is the exact opposite of some Ruddlesden--Popper-type layered perovskites.
  The \hbox{(B site metal)--O$_6$} octahedra in layered perovskite oxides are often distorted by the Jahn--Teller effect
  \cite{1999SIM,2017DL_NHL}. 
  The F atoms selectively occupy the apical sites and lead to significantly longer 
  \hbox{(B site metal)--F$^{\mathrm{apical}}$} distance than \hbox{(B site metal)--O$^{\mathrm{apical}}$} 
  and O$^{\mathrm{equatorial}}$ distances \cite{1994AI,1995NEE,1999SIM,1996NEE,2001HEC,2016SU}.
  However, the oxyfluoride {\pbbody} is less distorted than the isostructural oxide {\bibody}, as shown in Figure~\ref{fig.strExpt}. 
}

\vspace{2mm}
\tadd{
  Table \ref{tab.strparams} compares the experimental structure
  and theoretical F-in-site6 and F-in-site5 structures of {\pbbody}.
  The theoretical values are given as the mean values of the three functionals.
  The errors indicate the unbiased deviations of the means. 
  The F-in-site5 is slightly less symmetric than F-in-site6, but their differences are not significant, which 
  indicates that the low-symmetric anionic configurations around the cations in {\pbbody} 
  are barely due to the steric effects of 6$s^2$ lone pairs: 
  the structure is naively low-symmetric. 
  The less symmetric structure of {\bibody} than {\pbbody} can be simply explained by that
  Bi$^{3+}$ cations are more positive than Pb$^{2+}$ cations and prefer to be closer to anions.  
  In the case of {\bibody}, the steric effects of 6s$^2$ lone pairs may also contribute
  to the low-symmetric anionic configurations as discussed based on the MEM analyses in the SI.
  The finding from Table \ref{tab.strparams} demonstrates that the influence of 6$s^2$ lone pairs
  on the structural distortion might be similarly not significant 
  in some of the other Pb--based oxyfluorides
  such as Pb$_2$Ti$_2$O$_{5.4}$F$_{1.2}$ \cite{2016OKA} and Pb$_2$OF$_2$ \cite{2019INA}.    
}

\begin{table}[htbp]
  \begin{center}
    \caption{
      \tadd{
        \label{tab.strparams}\ghost{tab.strparams}
        \tadd{
          Parameters of the experimental structure and DFT structures (F-in-site5 and F-in-site6) obtained for {\pbbody}.
          The values for the DFT structures are given as the mean values of the three functionals.
          The errors indicate the unbiased deviations of the means. 
          The second rows for the F-in-site6 and F-in-site5 indicate the differences from experimental values in percentages.
        }
      }
    }
    \scalebox{0.9}{
      \begin{tabular}{cccccccccc}
        \hline
        & & $a$ & $b$ & $c$ & $\beta$ ($^\circ$) & Pb-O/F4 & Pb-O/F5 & Pb-O/F6$^{\mathrm{shorter}}$ & Pb-O/F6$^{\mathrm{longer}}$ \\
        \hline
        Expt. & $\AA$ & 14.63 & 3.83 & 10.75 & 135.58 & 2.56 & 2.95 & 2.42 & 2.56 \\
        \hdashline       
        \multirow{2}{*}{F-in-site6} & $\AA$  & 15.07(8) & 3.84(1) & 10.54(7) & 133.44(6) & 2.54(1) & 2.93(2) & 2.44(1) & 2.58(1) \\
        & \% & $+$3.0(6) & $+$0.3(3) & $-$2.0(7) & $-$1.6(1) & $-$0.1(3) & $-$0.8(5) & $+$1.0(2) & $+$0.9(2) \\
        \hdashline 
        \multirow{2}{*}{F-in-site5} & $\AA$ & 14.83(7) & 3.84(1) & 10.60(6) & 132.30(2) & 2.58(1) & 3.03(1) & 2.37(1) & 2.58(2) \\
        & \% & $+$1.4(5) & $+$0.2(3) & $-$1.4(6) & $-$2.4(1) & $+$1.6(3) & $+$2.7(4) & $-$2.0(4) & $+$0.9 (6) \\
        \hline
      \end{tabular}
    }
  \end{center}
\end{table}

\section{Conclusion}
\label{sec.conc} \ghost{sec.conc}
\tadd{
  A combination of $^{19}$F MAS NMR experiments and DFT simulations were used to investigate the anionic ordering in {\pbbody}. 
  The $^{19}$F MAS NMR experiments showed that F atoms predominantly occupy two of the six distinct available sites in \pbbody\, 
  in a ratio of 73:27.
  DFT calculations identified the majority and minority F occupation sites to be sites 6 and 5, respectively.
  The occupation ratios between sites 6 and 5 were quantitatively reproduced by theory, independent of the choice of functional.
  PDOS and COHP analyses revealed that the 6$s^2$ lone pairs of Pb atom may play a role ($\sim$0.1~eV/f.u.) in determining the site 6 (5) the majority (minority) site, against what is predicted by Pauling's second rule \cite{1960LP}.

  \vspace{2mm}
  The low-symmetric anionic coordinates around the cations in {\pbbody} have been 
  considered to be the consequence of the Jahn--Teller distortions by the 6$s^2$ lone pairs, 
  as well as {\bibody} \cite{1995KAH},  Pb$_2$Ti$_2$O$_{5.4}$F$_{1.2}$\cite{2016OKA}, 
  and some Aurivillius oxides ABi$_2$Nb(Ta)$_2$O$_9$ \cite{1996I_M}. 
  However, our DFT results indicate that the 6$s^2$ lone pairs may barely induce 
  a structural distortion in {\pbbody}. 
  This finding demonstrates that the influence of 6$s^2$ lone pairs on the structural distortion 
  might be similarly not significant in some of the other Pb--based oxyfluorides
  such as Pb$_2$Ti$_2$O$_{5.4}$F$_{1.2}$\cite{2016OKA} and Pb\tadd{$_2$}OF$_2$\cite{2019INA}.
}

\section*{\tadd{Author Contributions}}
\tadd{
K.O., T.I., and D.K. conceived the ideas of this work.
All the authors contributed to the discussion and writing of the paper.
K.O. and K.Y. synthesized the {\pbbody} and {\bibody} samples, 
analyzed the crystal structures, and performed MEM analyses.
Y.N., Y.T., and N.N. performed the $^{19}$F MAS NMR measurements.
T.I., D.K., and K.H. performed the ab initio calculations.
M.I., R.M., and F.A.R. supervised the work.
}

\section*{Acknowledgments}
\tadd{
  The authors acknowledge E. Heinrich for valuable help with manuscript preparation.
  The VESTA program \cite{2011MOM} was used to visualize
  the experimental and DFT crystal structures and electronic densities.
  The authors thank Prof. Sudo and Mr. Kitao of Kindai University for their help in the solid state NMR measurement.
}
The synchrotron radiation experiments were performed at the BL02B2 of SPring-8 with the approval
of the Japan Synchrotron Radiation Research Institute (JASRI) (Proposal No.~2016A1157 and No.~2018A1227).
The ab initio calculations were performed with the computational resources 
of the Research Center for Advanced Computing Infrastructure (RCACI) at JAIST. \tadd {The authors gratefully thank Division of Joint Research Center, Kindai University, for the 
solid-state NMR measurements.}
T.I. and F.A.R. acknowledge support from US Department of Energy, Office of Science, Basic Energy Sciences,
Materials Sciences and Engineering Division.
This work was also partially supported by a Grant-in-Aid for Scientific
Research on Innovative Area “Mixed Anion (Project, JP17H05489, JP19H04706)” (JSPS),
and a Grant-in-Aids for Scientific Research (C)
(Project JP16K05731 and 21K04659).
K.O. is grateful for financial support from Kansai Research Foundation for technology promotion.
K.H. is grateful for financial support from 
MEXT-KAKENHI (JP16H06439, JP19K05029, JP19H05169, and JP21K03400), 
and the Air Force Office of Scientific Research 
(Award Numbers: FA2386-20-1-4036). 
R.M. is grateful for financial supports from 
MEXT-KAKENHI (21K03400 and 19H04692), 
from the Air Force Office of Scientific Research 
(AFOSR-AOARD/FA2386-17-1-4049;FA2386-19-1-4015), 
and from JSPS Bilateral Joint Projects (with India DST).

\bibliography{references}

\begin{figure}[p]
  \centering
  \includegraphics[width=\hsize]{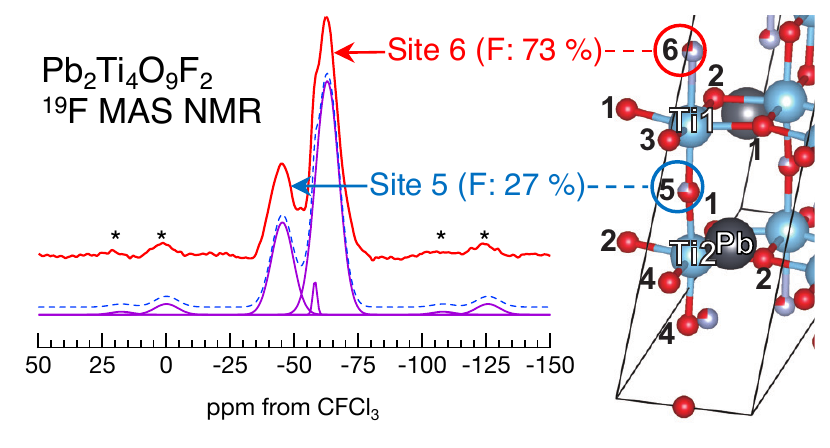}
  \caption{
    For Table of Contents Only.
  }      
\end{figure}
\end{document}